\documentstyle[aps,pra]{revtex}
\draft
\begin{document}

\twocolumn[\hsize\textwidth\columnwidth\hsize
\csname @twocolumnfalse\endcsname

\title{Quantum evolution according to real clocks}
\author{I\~{n}igo L. Egusquiza$^1$, Luis J. Garay$^2$,
and Jos\'{e} M. Raya$^{3}$}
\address{$^1$Fisika Teorikoaren Saila,
Euskal Herriko Unibertsitatea,
644 P.K., 48080 Bilbao, Spain\\
$^2$Instituto de Matem\'{a}ticas y F\'{\i}sica Fundamental,
CSIC, C/ Serrano 121, 28006 Madrid, Spain\\
$^3$Instituto de Astrof\'{\i}sica de
Andaluc\'{\i}a, CSIC,
Camino Bajo de Hu\'{e}tor, 18080 Granada, Spain}
\date{November 3, 1998}

\maketitle

\begin{abstract}

We characterize good clocks, which are naturally subject to
fluctuations, in statistical terms. We also obtain the master
equation that governs the evolution of quantum systems according
to these clocks and find its general solution. This master
equation is diffusive and produces loss of coherence. Moreover,
real clocks can be described in terms of effective interactions
that are nonlocal in time. Alternatively, they can be modeled by
an effective thermal bath coupled to the system.
\end{abstract}

\pacs{PACS: 03.65.Bz, 05.40.+j, 42.50.Lc
\hfill {EHU-FT/9807, quant-ph/9811009}}
]

\section{Introduction}

In quantum mechanics, spatial positions are described by quantum
observables. However, the situation with respect to the time
parameter is rather more involved, as has been known since Pauli
pointed out that there is no self-adjoint operator canonically
conjugate to the total energy, if the spectrum of the latter is
bounded from below \cite{pau}. Some attempts have been made to
circumvent this problem by making use of the concept of time of
arrival \cite{mug98}. Recent developments, however, have mapped
the problem of constructing a good time of arrival operator into
the problem of constructing a good time operator \cite{unruh},
thus showing that the special status of time is deeply embedded
into the structure of quantum mechanics. We are then bound to
use real physical clocks and rely on their readouts when
measuring the evolution of a quantum system.

Any real clock is inevitably subject to quantum fluctuations,
which introduce uncertainties in the equations of motion. For
instance, it has been shown that the finite mass and size of the
clock impose limitations in the measurement of spacetime
distances in the framework of general relativity \cite{wig}.
Some considerations have also been made on the role of quantum
clocks in the context of quantum cosmology
\cite{hart,unruhwald}. Simple models for quantum clocks have
been proposed, and the quantum evolution of a system according
to a quantum clock suitably coupled with it has also been
studied \cite{wig,hart,unruhwald,ash,page,ahar}. The general
conclusion is that the system becomes more and more perturbed as
the resolution of the clock is improved. Even more, quantum
gravity may well imply the existence of an absolute limit, the
Planck scale, to the accuracy of spacetime-distance measurements
and, in particular, to clock synchronization (for a review, see,
e.g., Ref. \cite{95gar01}), with possible effects in the
low-energy regime \cite{qfoam}.

It follows that any quantum clock that we could possibly build
would lead to uncertainties and errors. These quantum errors,
however, are not the only source of randomness in the measure of
time. Real clocks are also subject to classical imperfections,
small errors, that can only be dealt with statistically. For
instance, an unavoidable classical source of stochasticity is
temperature, which will introduce thermal fluctuations in the
behavior of real clocks. Although this is not necessarily the
most important source of errors in modern-day atomic clocks, it
is nonetheless always present to some extent. In other words,
the third law of thermodynamics forbids the existence of ideal
clocks.

We will study, within the context of the standard quantum
theory, the evolution of an arbitrary system according to a real
clock. Unlike other works
\cite{wig,hart,unruhwald,ash,page,ahar}, we are not concerned
about the quantum dynamical behavior of the clock but only with
the readings of time that it provides. As stated before, these
readings will undergo errors, which will be described by a
stochastic process. It is in this sense that we regard these
clocks as classical, although this does not preclude a possible
quantum mechanical origin for the stochastic fluctuations.
Notice also that we will mostly consider clocks which are
decoupled from the system under study. In what follows we shall
not delve further into the source of stochasticity, but assume a
phenomenological description of it.

Let us be more specific about the meaning of this randomness in
the readings of the clock. Imagine a large ensemble of identical
systems, prepare one of them in a given initial state at initial
clock time $t=0$, and then measure the state of that system at
clock time $t$. If we repeat this procedure for all the systems
in the ensemble, the result will be a probability distribution
for the possible outcomes, its dispersion partially being a
consequence of the lack of knowledge of the precise ideal time
that has elapsed. Therefore, the evolution according to the
readouts of the real clock is nonunitary. In other words, the
use of real clocks induces loss of coherence in most physical
quantum states, as we will explicitly show.

To attain this objective we will start by analyzing in some
detail the characteristics that should be expected from a good
real clock. This is expounded in the next section. The
implications of the randomness in the readings of a real clock
for the quantum evolution of a system is explored in Sec.
\ref{sec:master}. In Sec. \ref{sec:effective}, we discuss the
evolution according to real clocks from an effective point of
view in terms of nonlocal interactions in time and also in terms
of a thermal bath coupled to the system under study. Decoherence
owing to clock errors is investigated in Sec.
\ref{sec:decoherence}. Finally, we summarize and conclude in the
last section with some general remarks.

\section{Good clocks}

Let us consider the phase space of a classical physical system,
divided in a set of ordered cells, such that the expected
evolution of the system moves the state from one cell $k$ to the
next $k+1$ in a time $\varepsilon$ approximately. An observer
keeps tally of which is the cell of largest $k$ that has been
visited, and thus the passage of time is recorded. Let $s_k$
denote the ideal time at which the readout of the clock is
$k\varepsilon$, i.e., $s_k$ is the ideal time of first arrival
of the system at the $k$th cell. For an ideal clock, the time
$s_k$ and the clock time $k\varepsilon$ would coincide. However,
for a real clock, these two quantities will differ by an error
$\Delta_k=s_k-k\varepsilon$, where it should be noticed that the
index $k$ pertains to the readout $k\varepsilon$ of the clock,
i.e., to the $k$th tick, and not to a preset ideal time. Given
any real discrete clock, its characteristics will be encoded in
the probability distribution for the stochastic sequence
\cite{libros} of clock errors, $P(\{\Delta_{k}\})$, which must
satisfy appropriate conditions, so that it describes a good
clock.

A first property is that Galilean causality should be preserved,
i.e., that causally related events should always be properly
ordered in clock time as well, which implies that $s_{k+1}>s_{k}$
for every $k$. In terms of the discrete derivative
$\alpha_{k}=(\Delta_{k+1}-\Delta_{k})/\varepsilon$ of the
stochastic sequence $\{\Delta_k\}$, we can state this condition
as requiring that, for any realization of the stochastic
sequence, $\alpha_{k}>-1$.

A second condition that we would require good clocks to fulfill
is that the expectation value of relative errors, determined by
the stochastic sequence $\{\alpha_{k}\}$, be zero, i.e.,
$\langle\alpha_{k}\rangle=0$ for all $k$. If this were not the
case, the clock would either systematically go fast or slow down,
and a redefinition through this systematic drift would provide us
with a well-centered clock. Consequently, the expectation value
for the absolute errors $\{\Delta_{k}\}$ will be constant.
Furthermore, since $k=0$ will be the time at which the systems
whose evolution we are studying are prepared, $\Delta_{0}$ will
not be stochastic and, without loss of generality, will be set to
zero by a simple translation of the origin of time, so that
$\langle\Delta_{k}\rangle=0$.

Another expectation that we would have for a good clock is that
it should always behave in the same way (in a statistical
sense). The difference between the ideal time for the $(k+1)$th
tick and the ideal time for the $k$th tick must be always the
same, statistically speaking, even if the actual errors
$\Delta_{k+1}$ and $\Delta_{k}$ are large by some statistical
fluctuation. Therefore, we can say that the clock behaves
consistently in time as a good one if those relative errors
$\{\alpha_{k}\}$ are statistically stationary, i.e., the
probability distribution ${\cal P}(\{\alpha_k\})$ for the
sequence of relative errors $\{\alpha_{k}\}$ (which can be
obtained from $P(\{\Delta_k\})$, and vice versa) must not be
affected by global shifts $k\to k+k_0$ of the readout of the
clock. Note that the stochastic process $\Delta_k$ need not be
stationary, despite the stationarity of the process $\alpha_k$.

It is also intuitively obvious that the one-point probability
distribution function for the variables $\alpha_{k}$ should be
highly concentrated around the zero mean, if the clock is to
behave nicely. Even more, it is to be expected for clocks with
small errors that all the higher-order cumulants be much smaller
than the correlation, which, in turn, should also be
bounded by a small number, i.e.,
$$
\langle\alpha_{k}\rangle=0\,,
\qquad\langle\alpha_{k}\alpha_{k-j}\rangle\equiv
c_{j}\leq c_0\ll 1 \,,
$$
where $c_j=c_{-j}$. The correlation for the sequence of absolute
errors $\{\Delta_k\}$ can then be easily obtained and has the
form $ \langle\Delta_k\Delta_l\rangle=\varepsilon^2
\sum_{i=1}^{k}\sum_{j=1}^{l}c_{i-j} $.

The correlation time $\vartheta$ for the stochastic sequence
$\{\alpha_k\}$ is given by
$\vartheta=\varepsilon\sum_{j=-\infty}^{+\infty} c_j/(2c_0)$. We
will introduce a new parameter $\kappa$ with dimensions of time,
defined as $\kappa^2=c_0\vartheta^2$. This comes about because,
when the errors of the clock have a thermal origin, $\kappa^2$ is
proportional to the temperature, and independent of $\vartheta$.
In general, the good-clock conditions imply
$\kappa\ll\vartheta$. As we shall see, $\vartheta$ cannot be
arbitrarily large, and, therefore, the ideal clock limit is given
by $\kappa\to0$.

Until now we have discussed general properties that a good clock
must fulfill, regardless of the physical system under study. In
addition to these properties, a good clock must have enough
precision in order to measure the evolution of the specific
system, which imposes further restrictions on the clock. On the
one hand, the characteristic evolution time $\zeta$ of the system
must be much larger than the correlation time $\vartheta$ of the
clock. On the other hand, the leading term in the asymptotic
expansion of the variance $\langle\Delta_k^2\rangle$ for large
$k$ is of the form $\kappa^2 (k\varepsilon/\vartheta)$ which
means that, after a certain period of time, the absolute errors
can be too large. The maximum admissible standard deviation in
$\Delta_k$ must be at most of the same order as $\zeta$. Then the
period of applicability of the clock to the system under study,
i.e., the period of clock time during which the errors of the
clock are smaller than the characteristic evolution time of the
system is approximately equal to $\zeta^2\vartheta/\kappa^2$. For
a good clock, $\kappa\ll\vartheta\ll\zeta$, as we have seen, so
that the period of applicability is much larger than the
characteristic evolution time $\zeta$.

Even though, so far, we have only spoken of good discrete clocks,
by analogy, we will consider continuous stochastic processes
$\alpha(t)$, with corresponding probability functionals ${\cal
P}[\alpha(t)]$. The conditions previously stated for the discrete
sequence $\{\alpha_k\}$ admit a straightforward generalization to
the continuous case. In what follows, we shall use the
formulation in the continuum.

\section{Master equation}
\label{sec:master}

We shall now obtain the evolution equation for the density matrix
of an arbitrary quantum system in terms of the clock time $t$.
Let $\rho_{\sc s}(s)$ be the density matrix for a quantum system
whose unitary evolution in the ideal Schr\"{o}dinger time
$s=t+\Delta(t)$ is provided by the time-independent Hamiltonian
$H$.

For any given realization of the stochastic process that
characterizes a good clock, and using the chain rule, we can
write the von Neumann evolution equation in terms of the clock
time $t$ as
$$
\partial_t\rho_{\sc s}(t+\Delta(t))=
-i(1+\alpha(t))\big[H,\rho_{\sc s}(t+\Delta(t))\big] \,,
$$
where $\hbar$ has been set to 1.

Let us now transform to the interaction picture in which the
density matrix has the form
$$
\rho^{\sc i}_{\sc s}(t+\Delta(t))=
e^{iHt}\rho_{\sc s}(t+\Delta(t))e^{-iHt} \,.
$$
Notice that the interaction term $\alpha(t)H$ has the same form
in both pictures because it is proportional to the free
Hamiltonian $H$.

Integrating the resulting equation between 0 and $t$, and
reintroducing the result for $\rho_{\sc s}^{\sc i}$, we obtain
the following integro-differential equation:
\begin{eqnarray}
&&\partial_t\rho^{\sc i}_{\sc s}(t+\Delta(t))=
-i\alpha(t)\big[H,\rho^{\sc i}_{\sc s}(0)\big]
\nonumber\\
&&-\int_{0}^t dt'\alpha(t)\alpha(t')
\big[H,\big[H,\rho^{\sc i}_{\sc s}(t'+\Delta(t'))\big]\big] \,.
\nonumber
\end{eqnarray}
In order to find the evolution equation in the time $t$, we have
to average this equation over all possible realizations
$\alpha(t)$ of the stochastic process with the functional weight
${\cal P}[\alpha(t)]$. The average of the density matrix
$\rho^{\sc i}_{\sc s}(t+\Delta(t))$ will be denoted by $\rho^{\sc
i}(t)$ and can be regarded as the density matrix of the system at
clock time $t$.

At the real time $t=0$, we impose the initial condition
$\rho^{\sc i}(0)=\rho^{\sc i}_{\sc s}(0)=\rho^{\sc i}_0$.
Additionally, for a good clock, $\langle\alpha(t)\rangle=0$, as
already discussed, and, as a consequence, the average of the
linear term in $\alpha(t)$ vanishes. Furthermore, the clock time
derivative $\partial_t$ and the average over $\alpha(t')$ commute
because ${\cal P}[\alpha(t')]$ is stationary. Finally, the
density matrix $\rho^{\sc i}_{\sc s}(t'+\Delta(t'))$ can be
expanded in powers of $\Delta(t')$. Then the average of the
integro-differential equation for the density matrix $\rho^{\sc
i}$ yields
$$
\dot\rho^{\sc i}(t)=-\int_{0}^{t} d\tau
c(\tau)\big[H,\big[H,\rho^{\sc i}(t-\tau)\big]\big]
+O(\langle\alpha^3\rangle) \,,
$$
where the overdot denotes derivative with respect to the clock
time $t$. We have also performed a change of the integration
variable from $t'$ to $\tau=t-t'$ and have introduced the
correlation function $c(\tau)$ for the stochastic process
$\alpha(t)$.

For a good clock, the higher order terms in $\alpha$ can be seen
to be much smaller than the $c(\tau)$ term by a factor
$(\kappa/\zeta)^2\ll1$, provided that the system evolves for a
time smaller than the period of applicability of the clock.
Since $\zeta\gg\vartheta$, the system does not evolve
significantly within a correlation time, and we can substitute
$\rho^{\sc i}(t-\tau)$ by $\rho^{\sc i}(t)$. This is the
so-called Markov approximation. The process $\Delta(t)$ will not
be Markovian in general and there is no reason for requiring
that the process $\alpha(t)$ has this property either. However,
and even though the Markov approximation refers to the system
and not to the clock itself, it renders the possible
non-Markovian character of the clock irrelevant. Furthermore,
for evolution times $t$ much larger than the correlation time
$\vartheta$, we can take the upper integration limit to
infinity.

The resulting master equation, once we go back to the
Schr\"{o}dinger picture, can be written as
$$
\dot\rho(t)=-i\big[H,\rho(t)\big]-(\kappa^2/\vartheta)
\big[H,\big[H,\rho(t)\big]\big] \,.
$$
Notice that, in the ideal clock limit, $\kappa\to0$, the unitary
von Neumann equation is recovered. We should also point out that
this master equation is not a truncation of the BBGKY hierarchy
\cite{huang}, and that irreversibility appears because the errors
of the clock cannot be eliminated once we have started using it.

Under the good-clock conditions, $\kappa\ll\vartheta$, we can
approximate ${\cal P}[\alpha(t)]$ by a stationary Gaussian
probability functional with zero mean and correlation given by
the correlation $c(t)$ of ${\cal P}[\alpha(t)]$. Although this
Gaussian approximation assigns a nonvanishing probability to
$\alpha(t)<-1$, this probability will be negligibly small since,
for good clocks, $c(t)\ll1$. Thus the Gaussian approximation to
good clocks fulfills the Galilean causality condition for all
practical purposes.

In the Gaussian approximation, there is essentially only one
good clock for which $\alpha(t)$ is Markovian, the
Ornstein-Uhlenbeck process \cite{libros}. In this case, the
correlation function for $\alpha(t)$ in the stationary regime is
$c(\tau)=(\kappa/\vartheta)^2 e^{-|\tau|/\vartheta} $. Since the
possible non-Markovian character of the clock does not influence
the time evolution of the system (provided that the condition
$\zeta\ll\vartheta$ is satisfied, as happens for good clocks),
the Ornstein-Uhlenbeck clock is generic in what concerns the
evolution of quantum systems according to real clocks.

\section{Effective descriptions}
\label{sec:effective}

The master equation corresponds to the evolution of a system
with a free Hamiltonian $H$ coupled with a classical noise
source $\alpha(t)$, with a probability functional distribution
${\cal P}[\alpha(t)]$, via the interaction Hamiltonian
$\alpha(t)H$. The path integrals for this system then follow the
pattern \cite{feyn}
$$
\int{\cal D}\alpha {\cal P}[\alpha(t')]\int{\cal D}q
{\cal D}p e^{i\left[\int dt(p\dot q-H)-
\int dt\alpha(t)H(t)\right]} \,.
$$
In the good-clock approximation, only the two-point correlation
function $c(\tau)$ is relevant, so that we can write the
probability functional as a Gaussian distribution. The
integration over $\alpha(t)$ is then easily performed to yield
$$
\int{\cal D}q{\cal D}p
e^{i\int dt(p\dot q-H)-\frac{1}{2}\int dtdt'c(t-t')
H(t)H(t')}\,.
$$
Therefore, we see that the effect of using real clocks for
studying the evolution of a quantum system is the appearance of
an effective interaction term in the action integral which is
bilocal in time. This can be understood as the first term in a
multilocal expansion, which corresponds to the weak-field
expansion of the probability functional around the Gaussian term.

This nonlocality in time admits a simple interpretation:
correlations between relative errors at different instants of
clock time can be understood as correlations between clock-time
flows at those clock instants. The clock-time flow of the system
is governed by the Hamiltonian and, therefore, the correlation
of relative errors induces an effective interaction term,
generically multilocal, that relates the Hamiltonians at
different clock instants.

The expression for the influence functional is, in the weak noise
limit, completely analogous to the expressions above.
Since the noise source is classical,
we see that there is no dissipative term there, nor in the master
equation \cite{feyn}. Moreover, as the interaction
term is proportional to $H$, there is no response of the system
to the outside noise, which means that the associated impedance
is infinite \cite{callen,91gar01}.

{}From a different point of view, the clock can be effectively
modeled by a thermal bath, with a temperature $T_{\rm b}$ to be
determined, coupled to the system. Let $H+H_{\rm int}+H_{\rm b}$
be the total Hamiltonian, where $H$ is the free Hamiltonian of
the system and $H_{\rm b}$ is the Hamiltonian of a bath that
will be represented by a collection of harmonic oscillators
\cite{91gar01}. The interaction Hamiltonian will be of the form
$H_{\rm int}=\xi H$, where the noise operator $\xi$ is given by
$$
\xi(t)=\frac{i}{\sqrt{2\pi}}\int_0^\infty d\omega
\chi(\omega)
[ a^{\dag}(\omega) e^{i\omega t}-a(\omega) e^{-i\omega t}]\,.
$$
In this expression, $a$ and $a^{\dag}$ are, respectively, the
annihilation and creation operators associated with the bath, and
$\chi(\omega)$ is a real function, to be determined, that
represents the coupling between the system and the bath for each
frequency $\omega$.

Identifying, in the classical noise limit, the classical
correlation function of the bath with $c(\tau)$, the suitable
coupling between the system and the bath is given by the spectral
density of fluctuations of the clock:
$$
{k_{\sc b}} T_{\rm b}\chi(\omega)^2=\int_{0}^\infty d\tau
c(\tau)\cos(\omega\tau)\,,
$$
where $k_{\sc b}$ is Boltzmann's constant. With this choice, the
master equation for evolution according to real clocks is
identical to the master equation for the system obtained by
tracing over the effective bath.

To go beyond the classical noise limit requires the introduction
of the usual quadratic dissipation term in the influence
functional \cite{feyn}. However, the peculiar coupling $\xi H$
implies that this term does not produce dissipation in the
equations of motion: the fluctuation-dissipation theorem, which
reflects the microscopic structure of the bath, is thus
fulfilled, but there is no dissipation.

\section{Decoherence}
\label{sec:decoherence}

The master equation contains a diffusion term and will therefore
lead to a loss of coherence \cite{nico}. However, this loss
depends on the initial state. In other words, there exists a
pointer basis \cite{zurek}, so that any density matrix which is
diagonal in this specific basis will not be affected by the
diffusion term, while any other will approach a diagonal density
matrix. The stochastic perturbation $\alpha(t)H$ is obviously
diagonal in the basis of eigenstates $\{|n\rangle\}$ of the
Hamiltonian, which is therefore the pointer basis: the
interaction term cannot induce any transition between different
energy levels $\omega_n$.

The components of the density matrix in this basis are
$\rho_{nm}=\langle n|\rho| m\rangle $. The master equation can be
solved exactly, its general solution being
$$
\rho_{nm}(t)= \rho_{nm}(0) e^{-
i\omega_{nm}t}e^{-(\omega_{nm})^2\kappa^2 t/\vartheta}\,,
$$
where $\omega_{nm}=\omega_{n}-\omega_{m}$. The smallest energy
difference $\omega$ provides the inverse of the characteristic
time for the evolution of the system, $\zeta=1/\omega$. The
smallest decay constant is $\omega^2\kappa^2/\vartheta$, equal to
the inverse of the period of applicability of the clock. By the
end of this period, the density matrix will have been reduced to
the diagonal terms and a much diminished remnant of those
off-diagonal terms with slow evolution. In any case, the von
Neumann entropy grows if the density matrix is not initially
diagonal in the energy basis.

The effect of decoherence due to errors of real clocks does not
only turn up in the quantum context. Consider for instance a
classical particle with a definite energy $E$ moving under a
time-independent Hamiltonian $H$. Because of the errors of the
clock, we cannot be positive about the location of the particle
in its trajectory on phase space at our clock time $t$. Therefore
we have an increasing spread in the coordinate and conjugate
momentum over the trajectory. For a generic system, this effect
is codified in the classical master equation
$$
\dot\varrho=\big\{H,\varrho\big\}+
(\kappa^2/\vartheta)\big\{H,\big\{H,\varrho\big\}\big\}\,,
$$
where $\varrho(t)$ is the probability distribution on phase space
in clock time. This classical master equation can be derived in a
manner completely analogous to the quantum one.

For simplicity, let us study the particular example of
one-dimensional Hamiltonian motion with closed orbits, with $H=
\omega J$, $\varphi$ being the angle variable with period $2\pi$
conjugate to the action variable $J$, and $\omega$ a constant
frequency characteristic of the system. The classical master
equation for the probability density $\varrho(\varphi,J;t)$ reads
$$
\partial_t\varrho=\omega\partial_\varphi\varrho+
(\omega^2\kappa^2/\vartheta)\partial_{\varphi}^2\varrho\,.
$$
This diffusion equation can be exactly solved by separation of
variables. The slowest decaying mode has, as before, a decay
constant $\omega^2\kappa^2/\vartheta$.

In the case of one particle that is released with energy $E$ and
initial angle $\varphi_{0}$, the probability distribution spreads
out over the corresponding connected component of the energy
shell, and tends to $\delta(J-E/\omega)/2\pi$ as clock time
grows. As we can see, the information about the $\varphi$
variable is washed out by the errors in our clock, which is
precisely the information that is not available in the quantum
case: if $J$ is completely known for a given quantum state, the
indeterminacy in its conjugate variable will be infinite, the
situation towards which the classical decoherence process tends.

Finally, it should be observed that the mechanism of decoherence
is neither tracing over degrees of freedom, nor coarse graining,
nor dephasing \cite{nico,alamos}. Even though there is no
integration over time introduced here by fiat, as happens in
dephasing in quantum mechanics, the spread in time due to the
errors of the clock has a similar effect, and produces
decoherence.

\section{Conclusions}

In our study of the evolution of quantum systems according to
real clocks, which are necessarily subject to errors, we have
first established a stochastic characterization of good real
clocks. Using this description, we have derived a master
equation for the quantum evolution in real-clock time and we
have also found its general solution in the basis of energy
eigenstates. The stochastic features of good real clocks and
their effects on the quantum evolution can be equivalently
described by means of interactions which are nonlocal in time.
They can also be effectively modeled by a quantum thermal bath.
The master equation exhibits a diffusion term which is
responsible for the loss of coherence of most initial states.
Finally, we have analyzed the evolution of classical systems
according to real clocks and reached analogous conclusions.

The third law of thermodynamics and the quantum fluctuations
prevent real clocks from being perfectly accurate. This suggests
that, strictly speaking, the Schr\"{o}dinger unitary evolution
equation is just an excelent approximation valid for
sufficiently short periods of time and that should be
substituted, along the lines proposed in this paper, by a
diffusive master equation in more general situations. This adds
a random aspect to the evolution of quantum systems. Indeed,
coherence is progressively lost until we reach the period of
applicability of the clock and, after that time,
unpredictability sets in, as we have seen. Even perfectly
isolated systems will suffer loss of coherence because of the
fluctuations of the real-clock and will appear as effectively
coupled to a reservoir.

\acknowledgments

We thank C. Barcel\'{o}, C. Cabrillo, P.F. Gonz\'{a}lez-D\'\i az,
G.A. Mena Marug\'{a}n and M.A. Valle Basagoiti for discussions.
J.M.R.  is also grateful to J.M. Quintana.
We had support from the University of the Basque Country, Project
UPV 063.310-EB225/95, from Junta de Andaluc\'{\i}a,
and from DGICYT (Spain), Projects PB94-0107 and PB93-0139.

\end{document}